# Parton Distribution Functions for Pseudoscalar Mesons in the Confining Effective Chiral Quark Theory

Parada Tobel Paraduan Hutauruk
*Department of Physics, Pukyong National University (PKNU), Busan 48513, Korea,*
paradatobelparaduan.hutauruk@gmail.com

Seung-il Nam
*Department of Physics, Pukyong National University (PKNU), Busan 48513, Korea,* sinam@pknu.ac.kr

# Parton Distribution Functions for Pseudoscalar Mesons in the Confining Effective Chiral Quark Theory

Parada Tobel Paraduan Hutauruk[1,2*] and Seung-il Nam[1,3]

1. Department of Physics, Pukyong National University (PKNU), Busan 48513, Korea
2. Department of Physics Education, Daegu University, Gyeongsan 38453, Korea
3. Center for Extreme Nuclear Matters (CENuM), Korea University, Seoul 02841, Korea

*E-mail: paradatobelparaduan.hutauruk@gmail.com; sinam@pknu.ac.kr

## Abstract

This review paper presents the gluon, sea, and valence quark distributions of the pseudoscalar (PS) mesons. The calculations of the parton structure of PS mesons are performed using the Bethe-Salpeter equation—Nambu-Jona-Lasinio (BSE-NJL) model, which offers a clear description of the dynamical chiral symmetry breaking (D$\chi$SB) of the low-energy non-perturbative quantum chromodynamics (QCD), with the help of the Schwinger proper-time regularization scheme that simulates the color QCD confinement. Our results for the dynamical quark mass—which emerged from the spontaneous chiral symmetry breaking (S$\chi$SB)—are generated via the chiral condensate in the chiral limit and beyond. Results for the valence parton distribution functions for the PS meson at the factorization scale $Q = 2 GeV$ are in excellent agreement with experimental data and the gluon distributions for the pion fitted well with the lattice QCD and Jefferson Lab Angular Momentum (JAM) QCD global fit analysis.

## Introduction

The internal structure of the Nambu-Goldstone bosons—such as pions and kaons—attracts significant attention from the particle physics community [1, 2]. As it is well known that pion and kaon have very simple structures—which contain the dressed quark and dressed antiquark ($q\bar{q}$), compared to the complex and complicated structure of the nucleon with three dressed quarks ($qqq$). Although the pion and kaon have a simple structure, in fact, we do not fully understand their quarks and gluons dynamics. The situation becomes more complicated because we do not have a pion or kaon target in the experiment. Therefore, we only have limited data for the pion and kaon [3, 4], compared to the nucleon data.

Information on the internal structure of the pion and kaon can be extracted through the parton distribution function (PDF), elastic form factor (EFF), parton distribution amplitude (PDA), fragmentation function (FF), and generalized parton distribution (GPD). However, these quantities could not be directly computed from quantum chromodynamics (QCD). To overcome this problem, we employ the QCD-inspired models—mimicking the QCD properties (spontaneously breaking of the chiral symmetry (SB$\chi$S) and color QCD confinement)—such as the Bethe-Slapeter equation—Nambu—Jona-Lasinio (BSE—NJL) model [5–16], Dyson-Schwinger equation (DSE) model [17–19], light front holography (LFH) model [20], statistical model [21], nonlocal chiral quark (NLChQ) model [22], and basic light-front quantization-NJL (BLFQ-NJL) model [23]. Besides these models, the lattice QCD simulation [24, 25]—which is built based on the first QCD principle technique—can be also used to calculate the above-mentioned quantities as well as the global QCD analysis, such as Jefferson Lab Angular Momentum (JAM) QCD analysis [26], xFitter QCD analysis [27], and NNPDF collaboration [28].

Based on the current global analysis and theoretical models, the prediction results among models and global analysis result in different power counting rules of the PDF for the pion. For instance, the JAM global analysis [26] has quite the same result as the BSE—NJL model [3] which predicts $(1 − x)^1$ and the DSE model predicts $(1 − x)^2$. Both predict different power counting rules at the endpoint $x$. To resolve these different prediction results, new data from modern experiment facilities are absolutely required [29–32].

Many attempts have been already made to investigate the internal structure of the pion and kaon both theoretically and experimentally and many impressive signs of progress have been achieved so far. For example, nowadays, many theoretical calculations are available in the literature for the PDF, EFF, FF, and GPD for the pion and kaon using the DSE model [15–17], BSE—NJL model [3–14], and other models. Also, some upgraded and modern experimental facilities will be built and planned to run in the near future to improve the kinematic coverage, such as the Jlab12 upgrade (JLab20), electron-ion collider (EIC) [29,30], electron-ion collider in China (EicC) [31], CERN-Super Proton Synchrotron (COMPASS++/AMBER) [32], and JPARC. These experimental facilities will provide more precise data with a wide coverage of kinematics, which is really needed to fully understand the internal structure of the pion and kaon (in general hadron structure) and the properties of QCD itself as an underlying theory of strong interaction.

In this review paper, we present our calculations for the PDFs of the pion and kaon to more fully understand their internal structure and the effect of the DχSB on the pion and kaon PDFs *via* mass generation. The calculations of the PDFs for the pion and kaon are performed using the BSE—NJL model—effective chiral quark theory (EχQT)—with the help of the proper-time regularization (PTR) scheme—that simulates the QCD confinement. We also calculate the BSE—NJL dynamical quark mass—clearly showing the current (bare) quark mass evolution due to its interaction with the vacuum.

The BSE—NJL model has been successfully applied and widely used in various physics phenomena, for example, the valence quark distribution functions [3], fragmentation function [4], transverse-momentum dependents [5], EFFs for the pion and kaon in the medium [6–7], and gluon distribution functions in the nuclear medium [11]—which may provide useful information for the gluon distribution for the finite nuclei and is the same in spirit to the previous study of the European Muon Collaboration (EMC) effect. However, the BSE—NJL model is built based on the local four-fermion (contact) interaction and no dynamics of the gluon and sea quarks—absorbing the coupling constant—in the BSE-NJL model's effective Lagrangian. To generate the gluon and sea-quark distributions for the pion and kaon at higher $Q^2$, we evolve the valence quark distributions within the pion and kaon via the Dokshitzer-Gribov-Lipatov-Altarelli-Parisi (DGLAP) QCD evolution [33] in the next-leading order (NLO). It is worth noting that the evolution of the PDFs at high-$Q^2$ is really needed in order to be able to compare with the experimental data.

The outline of this review paper is organized as follows. In Sec. II, we briefly review the SU(3) effective Lagrangian of the BSE-NJL model. We then provide the dynamical quark masses and other quark properties in the chiral limit and beyond. In Sec. III, we present the expression for the twist-2 distribution function for the PS-mesons. In Sec. IV, our numerical results and discussions are presented and the implications are discussed. Section V is devoted to the summary.

## BSE-NJL Model

This section briefly describes the effective Lagrangian of the BSE-NJL model, which is an effective chiral quark theory, and the properties of the quark as well as the kaon and pion. Note that the BSE-NJL model simultaneously maintains the important QCD properties, namely the SχSB and simulating confinement via the PTR scheme. To describe the kaon and pion, we introduce the SU(3) BSE-NJL Lagrangian—which can be written in terms of the local four-fermion (contact) interaction:

$$\mathcal{L}_{NJL} = \overline{q}\left(i\partial/-\hat{m}_q\right)q + G_\pi[(\overline{q}\lambda_a q)^2 + (\overline{q}\lambda_a \gamma_5 q)^2] - G_\rho[(\overline{q}\lambda_a \gamma^\mu q)^2 - (\overline{q}\lambda_a \gamma^\mu \gamma_5)] \quad (1)$$

where the quark fields are given by $q = (u, d, s)^T$ and $\hat{m}_q = diag(m_u, m_d, m_s)$ are the current (bare) quark mass matrix. $G_\pi$ and $G_\rho$ are the local four-fermion (contact) coupling constants, which have dimensions with the units of GeV$^{-2}$ and $\lambda_a$ are the Gell-Mann matrices in the flavor space with $\lambda_0 = \sqrt{\frac{2}{3}}\mathbb{1}_{3\times3}$. The standard solution to the NJL gap (dynamical) equation is written as $S_q^{-1}(p) = p/-M_q - i\varepsilon$—the dressed quark propagator is given in the quark flavor space by $S_q(p) = diag[S_u(p), S_d(p), S_s(p)]$. Thus the dynamical (dressed or constituent) quark mass—gap equation—in the PTR scheme is given by

$$M_q = m_q + \frac{3G_\pi M_q}{\pi^2}\int_{\tau_{UV}^2}^{\tau_{IR}^2}\frac{d\tau}{\tau^2}exp[-\tau M_q^2], \quad (2)$$

where $\tau_{IR}^2 = 1/\Lambda_{IR}^2$ is the infrared (IR) integration limit with the value of $\Lambda_{IR} = 240$ MeV, which is determined based on the QCD limit ($\simeq 200 - 300$ MeV) and $\tau_{UV}^2 = 1/\Lambda_{UV}^2$ stands for the ultraviolet (UV) integration limit of $\Lambda_{UV}$, which is determined by fitting to the pion weak-decay constant $f_\pi = 93$ MeV and pion mass $m_\pi = 140$ MeV. Note that to have a finite theory, the ultraviolet cutoff is employed in the PTR scheme to remove the poles $\tau = 0$ and $\Lambda_{IR}$ is applied to remove the particle propagation for a larger $\tau$ value [3–14].

In the BSE-NJL model, the pion and kaon are the relativistic bound states of dressed quark-dressed antiquark that can be evaluated by solving the Bethe-Salpeter equations (BSEs). In general, this BSE can be also applied to other types of mesons like vector mesons, axial-vector mesons, and light-heavy mesons. The BSEs solutions are given by the interaction channels of the two-body amplitude. For the pion and kaon, it is simply given by

$$t_{[\pi,K]}(p) = \frac{-2iG_\pi}{1+2G_\pi \Pi_{[\pi,K]}(p^2)},\qquad(3)$$

where the polarization insertions (bubble diagrams) for the pion and kaon are respectively expressed by

$$\Pi_\pi(p^2) = 6i\int \frac{d^4k}{(2\pi)^4} Tr[\gamma_5 S_u(k)\gamma_5 S_{\overline{d}}(k+p)],\quad(4)$$

$$\Pi_K(p^2) = 6i\int \frac{d^4k}{(2\pi)^4} Tr[\gamma_5 S_u(k)\gamma_5 S_{\overline{s}}(k+p)].\quad(5)$$

From the pole of the amplitude in Equation (3), we then evaluate the masses for the pion and kaon by solving the pole equations:

$$1 + 2G_\pi \Pi_\pi(p^2 = m_\pi^2) = 0,\qquad(6)$$

$$1 + 2G_\pi \Pi_K(p^2 = m_K^2) = 0.\qquad(7)$$

Taking the first derivative of the polarization insertions in Equation (4) and (5) with respect to $p^2$, it gives the meson-quark coupling constant, which is needed to calculate the PDFs for the pion and kaon. It has the form

$$g_{\pi q\overline{q}}^{-2} = -\frac{\partial \Pi_\pi(p^2)}{\partial p^2}\big|_{p^2=m_\pi^2}.\qquad(8)$$

with $m_\pi$ is the meson mass.

## Meson Parton Distribution

Here, we present the general expression for the leading-twist quark distributions for the pion and kaon, which is given by

$$q_m(x) = \frac{p^+}{2\pi}\int d\xi^- exp[ixp^+\xi^-]\langle[m]\mid \overline{q}(0)\gamma^+q(\xi^-)\mid [m]\rangle_c,\qquad(9)$$

where $x = \frac{k^+}{p^+}$ is the longitudinal momentum fraction of the parton for the mesons with $k^+$ is the plus component of the struck momentum of the quark, $p^+$ is the plus component of the meson (parent hadron) momentum, and $\xi$ is the skewness variable. The subscript $c$ stands for the connected matrix element. Based on Feynman diagrams of Reference [3], we similarly evaluate the quark and antiquark distributions for the pion and kaon are respectively given by

$$q_m(x) = ig_{mq\overline{q}}^2\int \frac{d^4k}{(2\pi)^4}\delta(k^+ - xp^+)Tr_{c,f,\gamma}$$
$$[\gamma_5\lambda_a^\dagger S_l(k)\gamma^+ P_{u/d}S_l(k)\gamma_5\lambda_a S_l(k-p)],\quad(10)$$

$$\overline{q}_m(x) = -ig_{mq\overline{q}}^2\int \frac{d^4k}{(2\pi)^4}\delta(k^+ + xp^+)Tr_{c,f,\gamma}$$
$$[\gamma_5\lambda_a S_l(k)\gamma^+ P_{\overline{d}/\overline{s}}S_l(k)\gamma_5\lambda_a^\dagger S_s(k+p)],\quad(11)$$

where the operators are defined respectively as

$$\hat{P}_{u/d} = \frac{1}{2}[\frac{2}{3}1_{3\times3}\pm\lambda_3+\frac{1}{\sqrt{3}}\lambda_8], \hat{P}_{\overline{d}/\overline{s}} = [\frac{1}{3}1_{3\times3}-\frac{1}{\sqrt{3}}\lambda_8].\qquad(12)$$

Note that the trace runs over the color, flavor, and Lorentz indices. Then the valence quark distributions for the mesons are evaluated through the moment, which has the form

$$\mathcal{A}_n = \int_0^1 dx x^{[n-1]}q_m(x),\qquad(13)$$

where $n$ is an integer. After performing the Ward-Takahashi-like identity (WTI) via $S(k)\gamma^+S(k) = -\partial S(k)/\partial k_+$, Feynman parameterization, and the PTR scheme, we finally obtain the valence quark and antiquark distribution functions for the mesons in the PTR scheme, which have the forms respectively

$$q_m(x) = \frac{3g_{mq\overline{q}}}{(4\pi)^2}\int_0^1 dx\int_{\tau_{UV}^2}^{\tau_{IR}^2}d\tau\left[\frac{1}{\tau}+x(1-x)(m_m^2-(M_u-M_{\overline{d}/\overline{s}})^2\right]exp[-\tau(x(x-1)m_m^2+xM_{\overline{s}/\overline{d}}^2+(1-x)M_u^2)],\qquad(14)$$

$$\overline{q}_m(x) = \frac{3g_{mq\overline{q}}}{(4\pi)^2}\int_0^1 dx\int_{\tau_{UV}^2}^{\tau_{IR}^2}d\tau\left[\frac{1}{\tau}+x(1-x)(m_m^2-(M_u-M_{\overline{d}/\overline{s}})^2\right]exp[-\tau(x(x-1)m_m^2+xM_u^2+(1-x)M_{\overline{s}/\overline{d}}^2)].\qquad(15)$$

The valence quark distributions for the mesons must guarantee to preserve the baryon number and momentum sum rules, which are defined respectively as

$$\int_0^1 dx[u_K(x)-\overline{u}(x)] = \int_0^1 dx[\overline{s}_K(x)-s_K(x)] = 1,\quad(16)$$

$$\int_0^1 dx x[u_K(x)+\overline{u}(x)+\overline{s}_K(x)+s_K(x)] = 1.\quad(17)$$

Now we turn to describe how to generate the gluon and sea-quark distribution from the QCD evolution, as we already mentioned before that the BSE-NJL model has no dynamics of the gluons and sea quarks. Through the NLO-DGLAP QCD evolution, the gluon and sea quark distributions are purely and dynamically generated. In the DGLAP QCD evolution, the valence quark distribution—known as the non-singlet (NS) quark distribution—is given by

$$q_{NS}(x) = q(x) - \overline{q}(x),\qquad(18)$$

where $q(x)$ and $\overline{q}(x)$ are the quark and antiquark distributions, respectively. The non-singlet quark distributions in the DGLAP QCD evolution are given by

$$\frac{\partial q_{NS}(x,Q^2)}{\partial ln(Q^2)} = \mathcal{P}_{qq}(x, \alpha_s(Q^2)) \otimes q_{NS}(x, Q^2), \quad (19)$$

where $\mathcal{P}_{qq}$ is the splitting function of the quark-quark—can be interpreted as the probability for the quark of type-$q$ with momentum fraction $z$ emitting the quark and becoming a new type of quark $q$ with momentum fraction $x$. The convolution product between the splitting function and the non-singlet quark distribution is defined by

$$\mathcal{P}_{qq} \otimes q_{NS} = \int_x^1 \frac{dz}{x} \mathcal{P}\left(\frac{x}{z}\right) q_{NS}(z, Q^2). \quad (20)$$

The singlet quark distribution, which is another type of quark distribution in the DGLAP QCD evolution, can be expressed by

$$q_s(x) = \sum_i q_i^+ = \sum_i q_i(x) + \overline{q}_i(x), \quad (21)$$

where $i$ represents the quark flavor. The singlet quark distributions are obtained by

$$\begin{bmatrix} \frac{\partial q_S(x,Q^2)}{\partial ln Q^2} \\ \frac{\partial g(x,Q^2)}{\partial ln Q^2} \end{bmatrix} = \begin{bmatrix} \mathcal{P}_{qq} & \mathcal{P}_{qg} \\ \mathcal{P}_{gq} & \mathcal{P}_{gg} \end{bmatrix} \otimes \begin{bmatrix} q_s(x, Q^2) \\ g(x, Q^2) \end{bmatrix}. \quad (22)$$

The gluon distributions for the mesons can be numerically calculated using Equation (21). Note that the splitting function can be expanded in terms of $\alpha_s(Q^2)$ in the perturbative region and it gives

$$\mathcal{P}(x, Q^2) = [\frac{\alpha}{2\pi}]\mathcal{P}^{(0)}(z) + [\frac{\alpha}{2\pi}]^2\mathcal{P}^{(1)}(z) + \cdots, (23)$$

where the first term is the leading order (LO) and the second term is the next-leading order (NLO). The NLO for the $\alpha_s$ is given by

$$\alpha_s = \frac{4\pi}{\beta_0} \frac{1}{ln[Q_A]} [1 - \frac{\beta_1}{\beta_0} \frac{lnln[Q_A]}{ln[Q_A]}] + \mathcal{O}(\frac{1}{ln^2[Q_A]}), \quad (24)$$

with $Q_A = Q^2/\Lambda_{QCD}^2$, $\beta_0 = \frac{11}{3}N_c - \frac{4}{3}N_f$, and $\beta_1 = \frac{34}{3}N_c^2 - \frac{10}{3}N_cN_f - 2C_FN_f$, where $N_c$ and $N_f$ are respectively the number of colors and flavors. The $\Lambda_{QCD}$ value depends on the number of flavors and the renormalization scheme.

## Numerical Results and Discussion

Here, results for the valence, gluon, and sea-quark distributions for the pion and kaon in comparison with the experimental data [1, 2], the lattice QCD simulation [23], and the JAM phenomenology global fit QCD analysis [24] are presented. To generate the gluon and sea-quark distributions for the pion and kaon and to compare the valence quark distribution with the experimental data, we have to evolve them at the factorization scale $Q^2 = 4GeV^2$ from the initial scale $Q_0^2 = 0.16GeV^2$. Results for the constituent quark mass as a function of the coupling of $G_\pi/G_{critical}$, for the chiral limit ($m_q = 0$) and beyond ($m_q \neq 0$). In the chiral limit, it shows that the dynamical quark mass originally emerges from the chiral quark condensate ($\langle q\overline{q}\rangle$). The dynamical quark mass increases when the interaction with the vacuum increases, as clearly shown in Figure 1. Beyond the chiral limit ($m_q \neq 0$), the dynamical quark mass is generated not only by the chiral quark condensate but also by the current quark mass, which is expected coming from the Higgs mechanism.

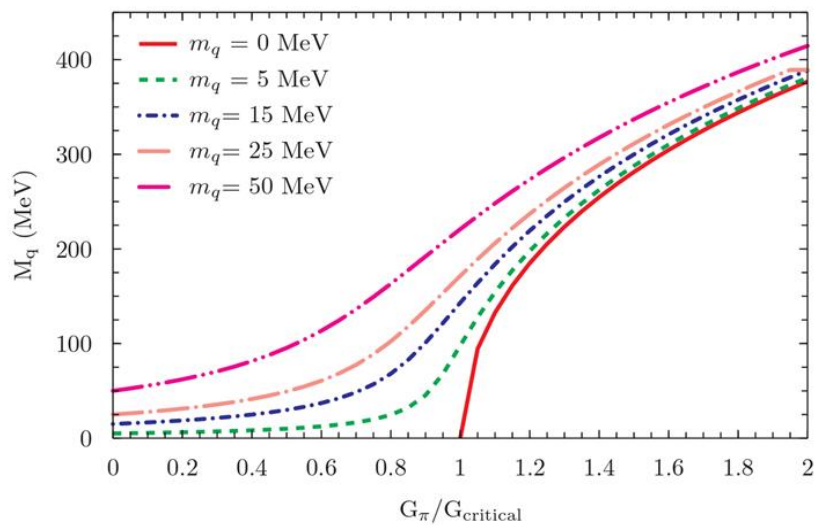

**Figure 1. The Constituent Quark Mass from the BSE-NJL Model**

Increasing the current quark mass will increase the constituent quark mass. However, in nature, the current quark mass for the up quark is quite small, which is around 5 MeV.

We now turn to show results for the PDFs for the pion at $Q^2 = 4 GeV^2$ are depicted in Figures 1(a)-(c). Figure 1(a) shows the valence quark distribution for the pion at $Q^2 = 4 GeV^2$—evolved via the NLO-DGLAP QCD evolution with the initial scale $Q_0^2 = 0.16 GeV^2$—in comparison with the experimental data [3].

It shows that our prediction results for the valence quark distribution for the pion have excellent agreement with the original and old experimental data [3] over the values of $x$. Also, this shows that the BSE-NJL model predicts $(1 − x)^1$ at $x \to 1$, which is consistent with the JAM results in Ref. [26]. However, our results are rather different from the reanalysis data [5]. In Figure 2(a), we show the gluon distribution for the pion in comparison with the JAM global QCD analysis [26] and lattice QCD simulation [25]. Our results indicate that the pion gluon distribution is in good agreement with the results in Refs. [25, 26], but, in small $x$, our result is rather different from the results of Refs. [25, 26]. Moreover, we also extract the sea-quark distribution for the pion—generated dynamically from the DGLAP equation and it then gives $S(x) = u^+(x) − u_v(x)$—as shown in Figure 2(c). Figure 2(c) shows that the sea quark carries very small light cone (LC) momentum in the large -$x$, but it is a little bit large in the low-$x$ ($x \leq 0.2$). Unfortunately, there is no experimental data to compare at the moment. Therefore, we have to wait for the data for the sea-quark distribution for the pion from modern experiment facilities.

Next, we also show our result for the kaon, which has a different type of quarks, as depicted in Figures 3(a)-(c).

Figure 3(a) shows the valence quark distribution for the kaon at $Q = 2 GeV$. The strange quark distribution for the kaon is larger than the up-quark distribution at $0.2 < x < 1.0$. However, the up-quark distribution for the kaon is larger than the strange-quark distribution at $x \leq 0.2$, where, in this region, the gluon and sea-quark distributions are expected to be larger, as seen in Figs. 3 (b) and (c). It indicates that the up-quark distribution in the small-$x$ carries more LC momentum than the strange-quark, then the strange quark carries more LC momentum, and the strange quark carries the most kaon momentum. Another momentum of the kaon is carried by the gluons and sea quarks. Interestingly, in the small-$x$, the up-quark carries more pion momentum than the strange quark. This means that the gluon and sea quark carries less inside the kaon in this region, in comparison with that in other regions.

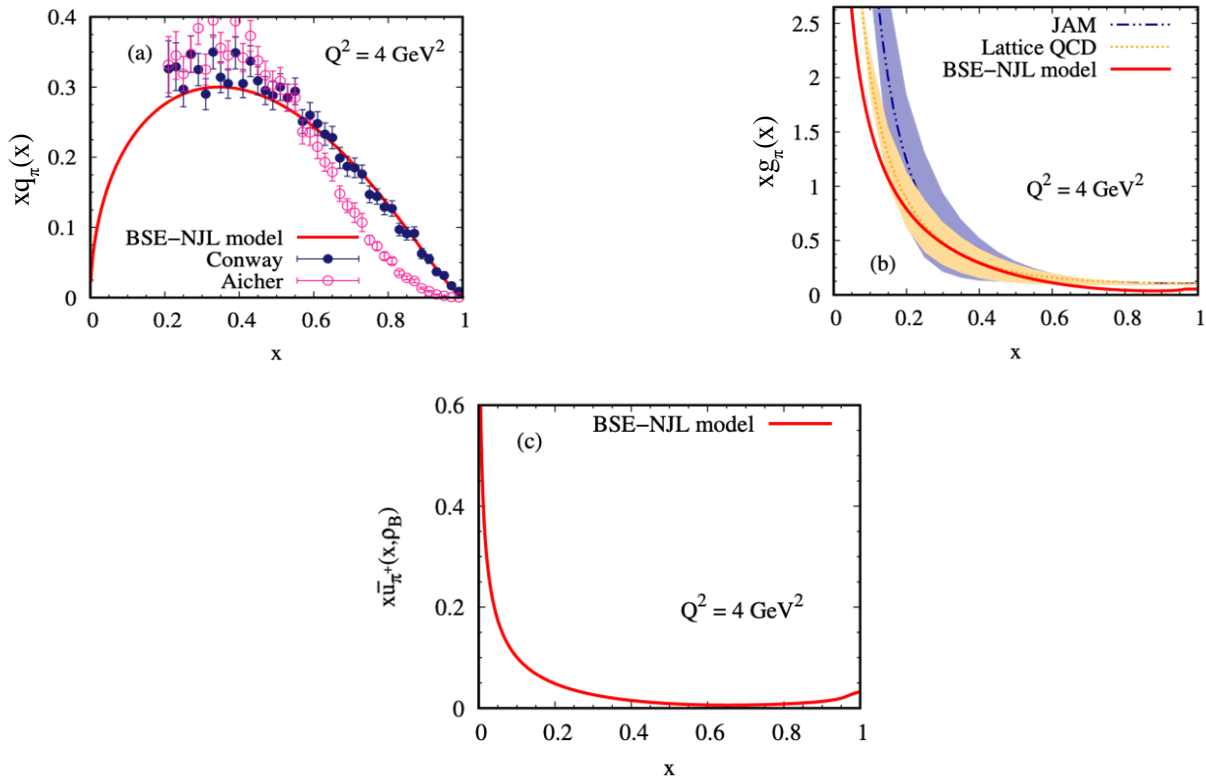

**Figure 2.** (a) Valence Quark Distributions for the Pion Calculated in the BSE-NJL Model at $Q = 2 GeV$ in Comparison with Experimental Data [3], (b) Gluon Distributions in Comparison with the JAM QCD Analysis [26] and Lattice QCD [25], and (c) Sea Quark Distributions

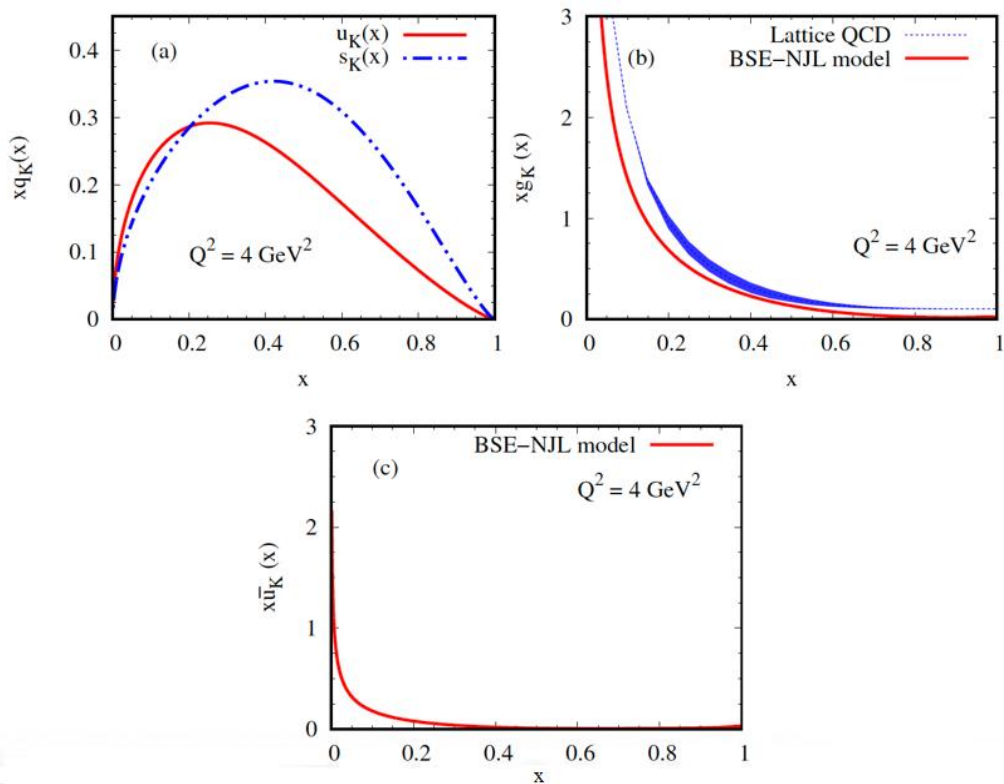

**Figure 3. The Same as in Figure 2, but for the Kaon**

Gluon distribution for the kaon is shown in Figure 3(b), in comparison with the lattice QCD result [35]. Our result for the gluon distribution for the kaon is not fitted well with the current lattice result. It is because the lattice calculation is very far from the pion's physical mass. Hence, comparing our results with the lattice data in the same pion physical mass could be more interesting and challenging.

Finally, we also calculate the sea-quark distribution for the kaon, which is similar to the pion, but we do not have available data at the moment for the sea-quark distribution for the kaon. So, we just leave our prediction and have to wait for the new experimental data and the lattice QCD result.

## Summary

As a summary, in this review paper, we have investigated the PDFs for the pion and kaon in the chiral limit and beyond using the BSE-NJL model with the help of the proper-time regularization scheme, simulating the QCD confinement. We employed the NLO-DGLAP QCD evolution to purely and dynamically generate the gluon and sea-quark distributions for the pion and kaon, which are absent in the initial model scale. We then compare our valence, gluon, and sea-quark distributions at $Q^2 = 4$ $GeV^2$ with the available experimental data, lattice QCD simulation result, and global QCD analysis.

We find that our results for the valence and gluon distributions for the pion and kaon at $Q^2 = 4$ $GeV^2$ are in excellent agreement with the available experimental data [3,4], the lattice QCD simulation [25,35], and JAM QCD analysis [26], respectively. Also, our results for the sea-quark distributions for the pion and kaon are presented.

More theoretical studies on the pion and kaon as well as other mesons using the more sophisticated models, which are the features of the models that exactly capture the QCD features, such as the DχSB and confinement, and the lattice QCD and QCD global analysis are really needed to better understand their internal structures, in particular, the gluon dynamics in the meson, which are potentially needed to interpret the coming future EIC data that will be available in the next 15 years. Moreover, new experimental data are absolutely required to resolve the remaining unsolved problems in the meson structure.

## Acknowledgment

This work was supported by the National Research Foundation of Korea (NRF) grants funded by the Korean government (MSIT) Nos. 2018R1A5A1025563, 2022R1A2C1003964, and 2022K2A9A1A06091176.